\documentclass[%
 amsmath,amssymb,
 reprint,%
 twocolumn
]{revtex4-1}

\usepackage{graphicx}
\usepackage{dcolumn}
\usepackage{bm}

\begin{document}

\preprint{AIP/123-QED}

\title{Phonon-assisted resonant tunneling of electrons in graphene-boron nitride transistors}

\author{E.E. Vdovin$^{1,5,6}$, A. Mishchenko$^2$, M.T. Greenaway$^1$, M.J. Zhu$^2$, D. Ghazaryan$^2$, A. Misra$^3$, Y. Cao$^4$, S. V. Morozov$^{5,6}$, O. Makarovsky$^1$, A. Patan\`e$^1$, G.J. Slotman$^7$, M.I. Katsnelson$^7$, A.K. Geim$^{2,4}$, K.S. Novoselov$^{2,3}$, L. Eaves$^1$}

\affiliation{$^1$School of Physics and Astronomy, University of Nottingham NG7 2RD UK \\
$^2$School of Physics and Astronomy, University of Manchester, Manchester, M13 9PL, UK \\
$^3$National Graphene Institute, University of Manchester, Manchester, M13 9PL, UK \\
$^4$Centre for Mesoscience and Nanotechnology, University of Manchester, M13 9PL, UK \\
$^5$Institute of Microelectronics Technology and High Purity Materials, RAS, Chernogolovka 142432, Russia \\
$^6$National University of Science and Technology "MISiS", 119049, Leninsky pr. 4, Moscow, Russia \\
$^7$Radboud University, Institute for Molecules and Materials, Heyendaalseweg 135, 6525 AJ Nijmegen, The Netherlands
}

\date{\today}

\begin{abstract}
We observe a series of sharp resonant features in the differential conductance of graphene-hexagonal boron nitride-graphene tunnel transistors over a wide range of bias voltages between ~10 and 200 mV. We attribute them to electron tunneling assisted by the emission of phonons of well-defined energy.  The bias voltages at which they occur are insensitive to the applied gate voltage and hence independent of the carrier densities in the graphene electrodes, so plasmonic effects can be ruled out.  The phonon energies corresponding to the resonances are compared with the lattice dispersion curves of graphene-boron nitride heterostructures and are close to peaks in the single phonon density of states.  
\end{abstract}

\maketitle

\begin{figure}[!t]
  \centering
\includegraphics[width=0.9\linewidth]{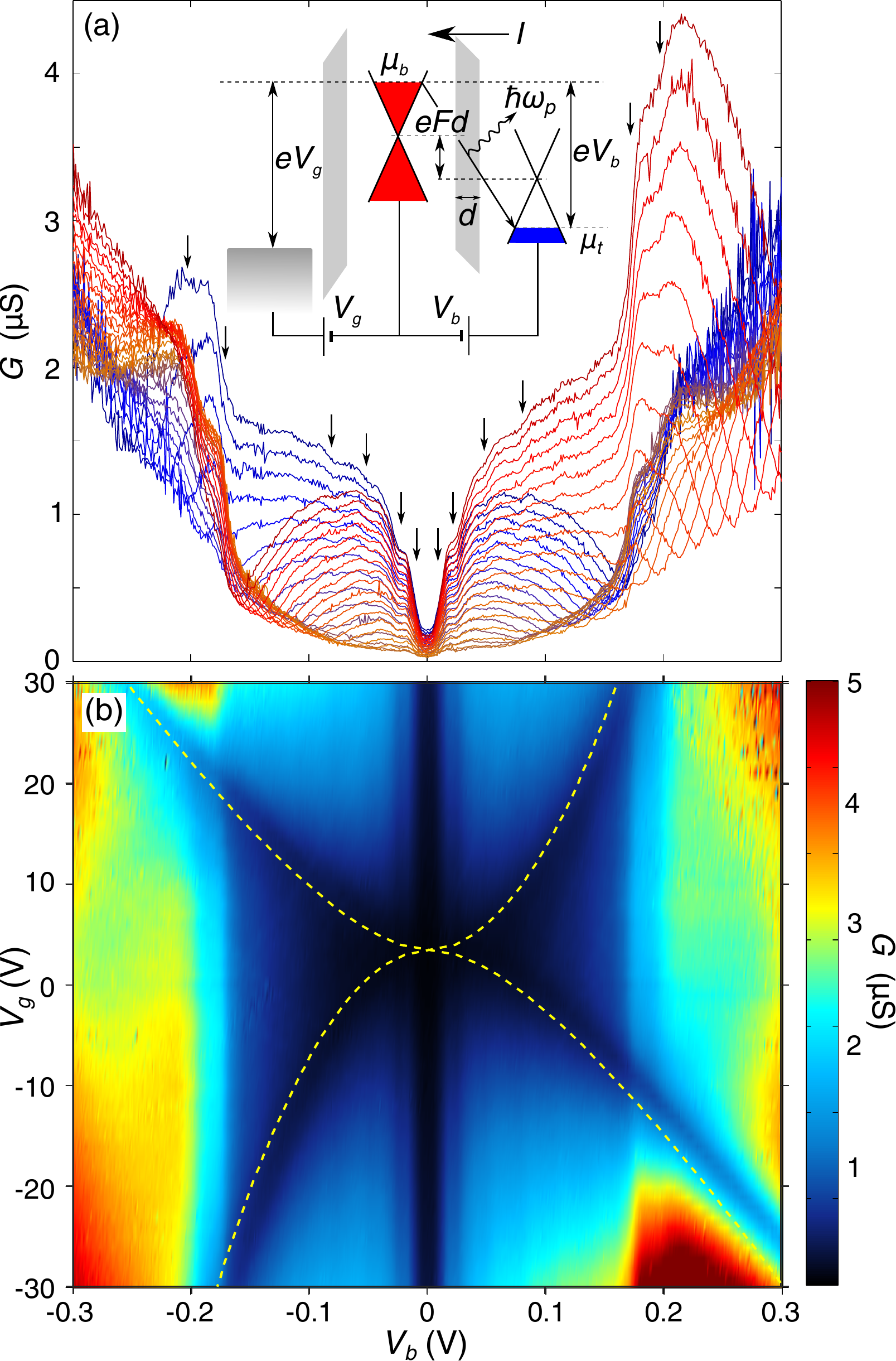}
  \caption{ Differential conductance of Device 1 at $T=4$ K.  (a) $G(V_b)$ for $V_g=-30$ V (red) to 30 V (blue) and intervals, $\Delta V_g = 2$ V.  Inset: schematic band diagram of Device 1 with bias, $V_b$ and gate, $V_g$, voltages applied to the monolayer graphene electrodes which are separated by an hBN barrier of thickness $d=0.9$ nm.  $\mu_{b,t}$ are the chemical potentials of the bottom (b) and top (t) electrodes and $F$ is the electric field across the barrier. A phonon assisted tunnel process is shown schematically. (b) Color map of $dI/dV_b$ for a range of $V_g$ and $V_b$. } 
\label{fig:1}
\end{figure}

The discovery of the remarkable electronic properties of graphene \cite{Novoselov2004,Novoselov2005} has been followed by an upsurge of interest in other layered materials such as hexagonal boron nitride (hBN), the transition metal dichalcogenides and the III-VI family of layered semiconductors.  These anisotropic layered materials have strong in-plane bonds of covalent character, whereas the inter-layer bonding arises from weaker van der Waals-like forces, so that crystalline flakes, one or a few atomic layers thick, can be exfoliated mechanically from bulk crystals.  These developments have led to the realisation of a new class of materials, van der Waals (vdW) heterostructures, in which nanoflakes of different materials are stacked together in an ordered way \cite{Dean2010,Geim2013,Ferrari2015}.  These heterostructures possess unique properties which can be exploited for novel device applications.  An example is a tunnel transistor in which a barrier is sandwiched between two graphene layers and mounted on the oxidised surface layer of a doped Si substrate \cite{Britnell2012,Feenstra2012}.  The tunnel current flowing between the two graphene layers can be controlled by applying a gate voltage to the doped Si layer and arises predominantly from resonant processes in which the energy, in-plane momentum and chirality of the tunneling electron are conserved \cite{Britnell2013a,Mishchenko2014,Greenaway2015}.  

Previous investigations of electron tunneling in a wide variety of metal-insulator diodes \cite{Chynoweth1962} and conventional semiconductor heterostructures \cite{Eaves1985} have demonstrated that electrons can tunnel inelastically, with the emission of one or multiple phonons.  Atomically-resolved scanning tunneling spectroscopy measurements on mechanically-cleaved graphene flakes with a tuneable back gate have revealed the presence of phonon-assisted tunneling \cite{Brar2007,Zhang2008,Wehling2008,Natterer2015}.  The multi-component nature of our vdW heterostructure gives rise to a more complex set of lattice dispersion curves than for graphene \cite{Slotman2014,Serrano2007,Maultzsch2004,Wirtz2004,Ferrari2006} and to phonon-assisted tunneling, as shown recently for a graphite-hBN-graphene transistor \cite{Jung2015}. An impetus to the study of electron-phonon interactions is the recent discovery of superconductivity in graphene-based vdW heterostructures \cite{Xue2012,Petrovic2013,Guzman2014,Chapman2015}. 

Here we investigate tunnel transistors in which a $\sim$1 nm layer of hBN is sandwiched between monolayer graphene electrodes.  We observe a series of sharp resonant peaks in the electrical conductance over a wide range of bias voltage, gate voltage and temperature.  This spectrum can be understood in terms of inelastic transitions whereby electrons tunneling through the hBN barrier emit phonons of different and well-defined energies between $\sim 12$ and 200 meV, covering the range of lattice phonon energies in these heterostructures.  The resonances correspond closely to van Hove-like peaks in the single phonon density of states of the heterostructure, with the strongest peaks arising from the emission of low and high energy optical mode phonons.

A schematic energy band diagram of our device and circuit is shown in the inset of Fig. \ref{fig:1}(a).  The bottom graphene layer is mounted on an atomically-flat hBN layer, placed above the silicon oxide substrate, and the active region of the device is capped with an hBN protective top layer.  The tunnel current, $I$, was measured as a function of the bias voltage, $V_b$, applied between the two graphene electrodes and the gate voltage, $V_g$, applied across the bottom graphene electrode and the doped Si gate electrode.   

Fig. \ref{fig:1}(a) shows plots of differential tunnel conductance, $G(V_b) = dI/dV_b$, measured at a temperature of $T = 4$ K.  The form of the $G(V_b)$ curves is strongly dependent on $V_g$.  Close to $V_b = 0$, $G \approx 0$ at all gate voltages.  With increasing $|V_b|$, the conductance increases in a series of well-defined steps, indicated by vertical arrows.  We attribute each step to inelastic phonon-assisted tunneling in which an electron emits a phonon and tunnels from close to the Fermi energy in one electrode to an empty state near the Fermi energy in the other electrode, with the emission of a phonon of well-defined energy, $\hbar \omega_p$.  Fig. \ref{fig:1}(b) shows a color map of $G(V_b,V_g)$ in which some of these step-like features are discernible as a series of faint vertical striped modulations in the color map.  Also visible is a dark blue, X-shaped region in which $G$ is small.  This corresponds to the passage of the chemical potential through the Dirac point of the two monolayer graphene electrodes as $V_b$ and $V_g$ are varied; here the conductance is suppressed due to the small density of electronic states into which electrons can tunnel.  Using an electrostatic model \cite{Britnell2012}, which includes a small amount of doping in the bottom electrode (p-type, $2.5\times10^{11}$ cm$^{-2}$) as a fitting parameter, we determine the condition for the intersection of the chemical potential with a Dirac point in each of the two graphene electrodes.  The calculated loci of these intersections are shown by the yellow dashed line in Fig. \ref{fig:1}(b); they correspond closely with the measured X-shaped low conductance region.

\begin{figure}[!t]
  \centering
\includegraphics[width=0.9\linewidth]{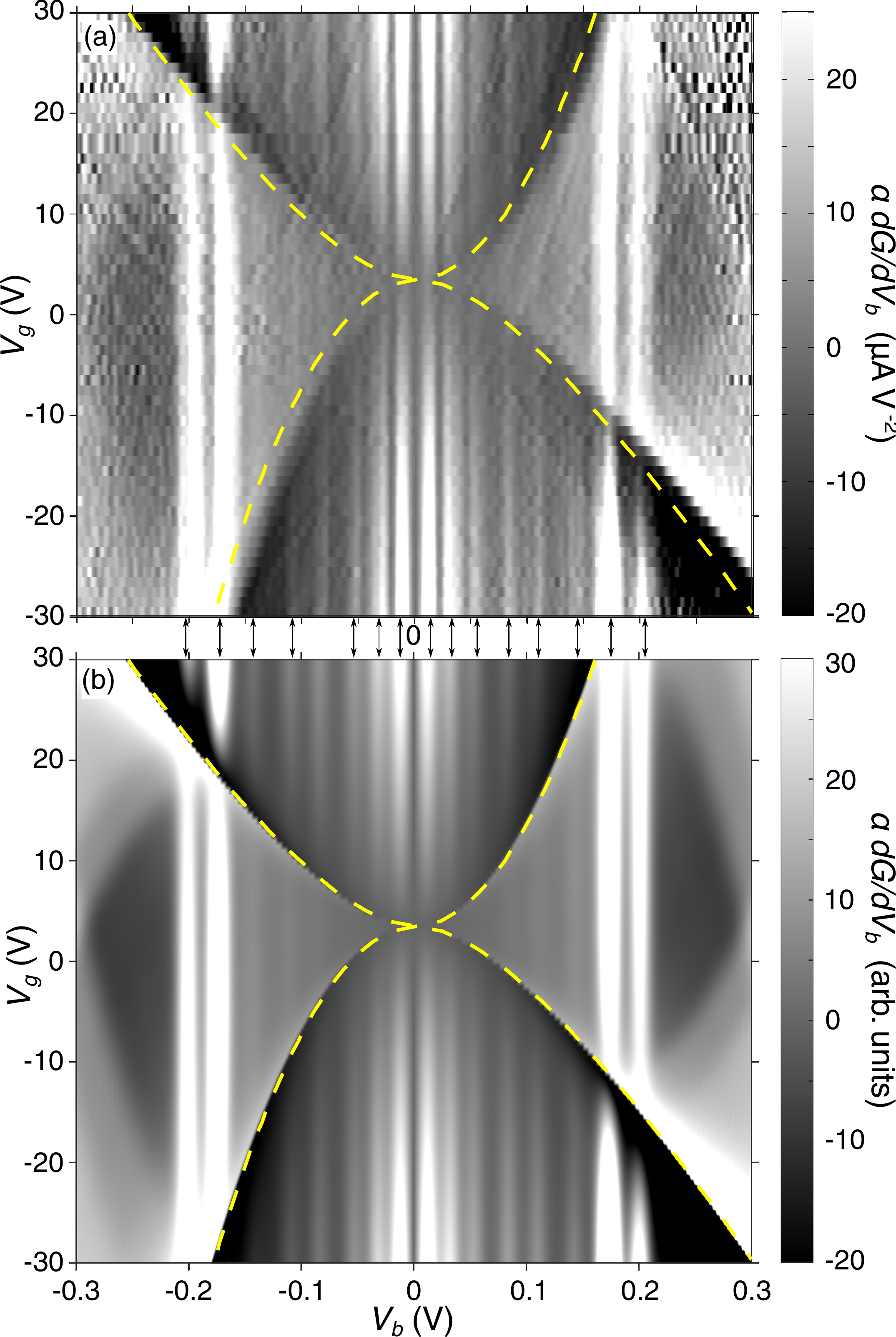}
  \caption{(a) Measured plots $\alpha dG/dV_b$ for Device 1 and (b) calculated grey scale map of $\alpha dG/dV_b$ for $T=4$ K, best fit to data in Fig. \ref{fig:2}(a) using the model described in the text and in Table 1.  Yellow dashed curves show when the chemical potential in a graphene layer intersects with the Dirac point in that layer.} 
\label{fig:2}
\end{figure}

The vertical stripes in the $G(V_b,V_g)$ map are faint because the step-like modulation in $G(V_b)$ is only a small fraction of the total conductance.  Most of the monotonic increase of $G(V_b)$ with $V_b$ can be partly eliminated by taking the second derivative, $dG / dV_b$, which reveals the weak but sharp phonon-assisted resonances more clearly. Fig. 2(a) shows a grey scale contour map of $\alpha dG / dV_b$, where $\alpha = |V_b|/V_b=\pm1$.  Here the phonon-assisted tunneling features appear as easily discernible bright vertical stripes, indicated by arrows, at well-defined values of $V_b$, at which $G(V_b)$ rapidly increases.  These values are independent of gate voltage but their amplitudes at low $V_b$ are significantly stronger at large positive and negative values of $V_g$.  We can exclude the possibility that the features are plasmon-related as the sheet density, $n$, in both graphene electrodes is strongly dependent on $V_g$: $n$ varies from $\sim 10^{12}$ cm$^{-2}$ (holes) through zero to $\sim 10^{12}$ cm$^{-2}$ (electrons) between $V_g=-30$ V and 30 V. Even though the plasma frequency of carriers in graphene varies relatively weakly with $n$ ($\sim n^{1/4}$), \cite{Hwang2007} plasmon-related resonances would have a significant gate voltage dependence which is not observed.
  
\begin{figure}[!t]
  \centering
\includegraphics[width=1.0\linewidth]{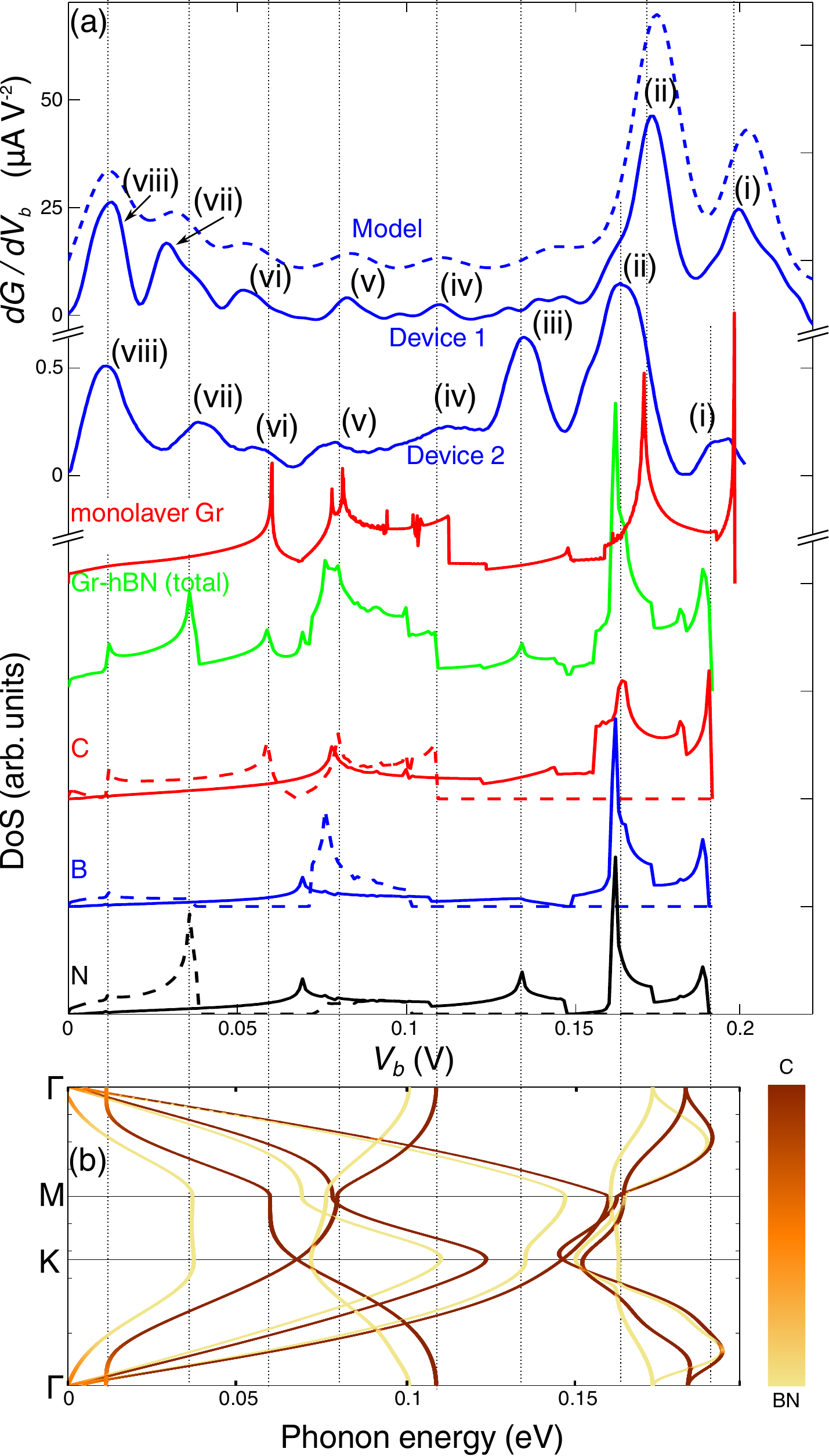}
  \caption{(a) The three top curves (blue): $dG/dV_b$ averaged over all gate voltages for our model (dashed) and measured data for Devices 1 and 2 (both solid).  Lower curves: total density of phonon states for monolayer graphene (red, monolayer Gr) and a graphene-hBN heterostructure (green, Gr-hBN (total)).  Lower three curves: the partial density of states associated with the carbon (red, C), boron (blue, B) and nitrogen (black, N) atoms of a graphene-hBN heterostructure.  Solid and dashed curves show contributions by in- and out of-plane phonons respectively.  (b) phonon dispersion of a graphene-hBN heterostructure \cite{Slotman2014}.  Vertical dotted curves are guides to the eye highlighting the alignments.}
\label{fig:3}
\end{figure}

Since the bias voltage values, $|V_b|$, of the weak resonant features are independent of gate voltage and are the same in both negative and positive bias, we can display them more clearly by averaging over all sixty of the measured $\alpha dG / dV_b$  plots between -30 V $< V_g <$ 30 V.  This procedure reduces significantly the level of background noise.  The result of this averaging procedure for Device 1 is shown in Fig. \ref{fig:3}(a).  It reveals the phonon-assisted resonances as a series of well defined peaks.  The corresponding plot for another device, Device 2, is also shown.  The overall forms of the two curves are qualitatively similar, with the exception of some notable differences e.g. the position of the strong peaks at high $V_b>0.12$ V.

To understand the physical origin of the peaks in Fig. \ref{fig:3}(a), we compare them to the one-phonon densities of states of monolayer graphene (red curve) and a graphene-hBN bilayer (green); the lower three curves show the partial density of states associated with the predominant motion of the carbon, boron and nitrogen atoms in the bilayer; the solid and dashed curves show contributions by in- and out of-plane phonons respectively. The full phonon dispersion curves of the graphene-hBN bilayer \cite{Slotman2014} are shown in Fig. \ref{fig:3}(b).  

The phonon density of states and the phonon dispersion curves were determined by using the ``phonopy'' package \cite{Togo2015} with the force constants obtained by the finite displacement method \cite{Kresse1995,Parlinkski1997}, using the Vienna ab initio simulation package (VASP) \cite{Kresse1996,Kresse1996prb}. 
For the phonon density of states a tetrahedron smearing was applied for higher accuracy. A detailed description of the computational methods can be found in ref. \cite{Slotman2014}.

At high bias, the two peaks labeled (i) and (ii) are close to the energies of the large densities of states associated with the weakly-dispersed, high energy optic phonons of monolayer graphene (Device 1) and a bilayer of graphene and hBN (Device 2).  Note that peak (iii) at 130 meV in Device 2 is absent in Device 1. This energy corresponds closely to the flat region of the dispersion curve of the graphene-hBN bilayer near the K-point of the Brillouin zone, which arises predominantly from vibrations of the nitrogen atom.  This difference, and the variation of the position of peaks (i) and (ii), between the two devices may arise from small differences in the relative lattice orientation of the graphene and hBN layers in the device. 

\begin{figure}[!t]
  \centering
\includegraphics[width=0.9\linewidth]{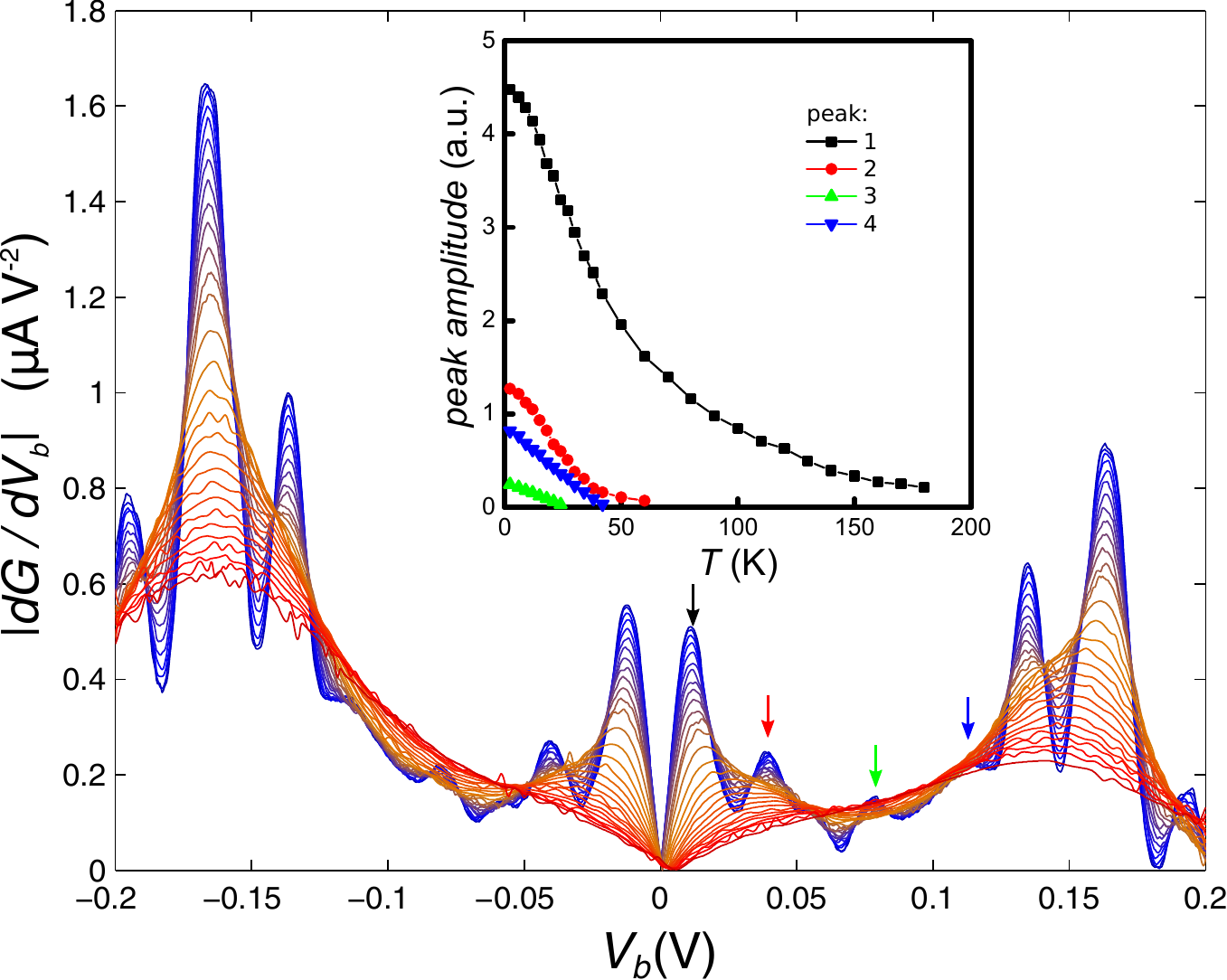}
  \caption{Temperature dependence of $|d^2I/dV_b^2|$ vs. $V_b$ in Device 2 measured from $T=2.3$ K to $T=180$ K (blue to red) for $V_g=40$ V.  Inset: peak amplitude vs. $T$, curve colors match peaks marked by correspondingly colored arrows in the main plot.}  
\label{fig:4}
\end{figure}

Both devices exhibit peaks around 110, 84 and 53 mV, labeled (iv), (v) and (vi), corresponding to prominent features in the calculated density of states plots and the flat regions of the dispersion curves.  An interesting feature of the data for both devices is the strong peak (viii) at low bias, around 12 mV, which we attribute to the weakly dispersed low-energy phonons close to the $\Gamma$-point of graphene-hBN.  This low-energy ``out-of-plane'' mode was intensively studied in inelastic X-ray spectroscopy measurement of bulk graphite and hBN \cite{Serrano2007}.  Note that the resonant peak (vii) observed at 32 mV can be associated with a peak in the phonon density of states of the graphene-hBN bilayer which arises predominantly from the motion of the nitrogen atoms and corresponds to the flat region of the lowest-energy acoustic mode at $\sim 30$ meV in the vicinity of the M- and K-points of the Brillouin zone. 

As shown in Fig. \ref{fig:4}, the resonant peaks broaden and their amplitudes decrease with increasing temperature, disappearing completely at temperatures $T \geq 150$ K.  This is consistent with the thermal broadening of the electron distribution functions around the Fermi energies of the two graphene electrodes.  Pauli blocking of electron tunneling for $eV_b < \hbar \omega_p$ between the graphene layers is diminished as more states become available with increasing thermal smearing around the Fermi energies.  

We fit the data in Fig. \ref{fig:2}(a), using a model in which an inelastic tunneling transition becomes allowed only when the difference between the chemical potentials, $\mu_b$ and $\mu_t$, in the bottom (b) and top (t) graphene layers respectively, exceeds $\hbar \omega_p^i$, which corresponds to the bias voltage of a particular phonon-assisted resonance peak, $i$, in the conductance.  At low temperatures (4 K), $eV_b = \mu_b-\mu_t-eFd$ greatly exceeds the smearing, $\sim 2 k_B T$, of the Fermi seas of the two graphene electrodes.  The emission of a phonon of energy, $\hbar \omega^i_p$, becomes possible when $eV_b$ exceeds $\hbar \omega^i_p$, thus opening up an inelastic scattering channel and giving rise to a step-like rise in the current and a resonant peak in $dG / dV_b$ when $eV_b = \hbar \omega^i_p$.  In our model the current is given by
\begin{widetext}
\begin{align}
I=\sum_iT(i) \int dE_b \int dE_t 
 D_b(E_b)D_t(E_t) \left\lbrace \Gamma(E_b-E_t-\hbar \omega^i_p) f_b(E_b)[1-f_t(E_t)] -  \Gamma(E_t-E_b-\hbar \omega^i_p) f_t(E_t)[1-f_b(E_b)] \label{eq:current} \right\rbrace 
\end{align}
\end{widetext}
where $E_{b,t}$ is the electron energy in the b and t layer, $D_{b,t}(E)$ is the density of electronic states in the b and t graphene layers (which are shifted energetically by $e F_b d$),  $\Gamma(E)=\exp(-E^2/2\gamma^2)$ with energy broadening $\gamma=5$ meV and $f_{b,t}$ is the Fermi function in the bottom and top electrodes. $T(i)$ is a relative weighting factor that depends on the electron-phonon coupling and phonon density of states for each inelastic transition.  We show the values of $T(i)$ used in our model in Table \ref{tab:1} which provides a qualitative indication of the relative strengths of the phonon-assisted processes.  

\begin{table}[!h]
\renewcommand{\arraystretch}{2}
\caption{Phonon energies, $\hbar \omega_p^i$, and weighting factors, $T(i)$, used in Eq. (\ref{eq:current}) to obtain the fit to the experimental data shown in Fig. \ref{fig:2}(b)  \label{tab:1}}
\begin{tabular}{ | c | c | c | c | c | c | c | c | c |}
\hline
  $i$ & 1 & 2 & 3 & 4 & 5 & 6 & 7 & 8 \\ \hline
  $\hbar \omega^i_p$ (meV) & 12 & 32 & 53 & 84 & 110 & 143 & 174 & 201 \\ \hline
  $T(i)$ & \ 1.0 \ & \ 0.58 \ & \ 0.30 \ & \ 0.26 \ & \ 0.24 \ & \ 0.23 \ & \ 3.53 \ & \ 1.81 \ \\  \hline
\end{tabular}
\end{table}

Using this model, and including phonon emission processes at threshold energies corresponding to the values of $V_b$, we obtain the grey-scale plot in Figure \ref{fig:2}(b), which simulates accurately with the measured data in Figure \ref{fig:2}(a).  In particular, the relative intensities of the vertical stripes are in good agreement with the measured data.  At high positive and negative $V_g$, the asymmetry in the measured intensities of the resonances for positive and negative $V_b$ is replicated by the model.  This confirms that the peaks arise from phonon-assisted tunneling of carriers from filled states near the chemical potential in one electrode into the empty states just above the chemical potential in the other.  

In conclusion,  our measurements reveal a rich spectrum of inelastic phonon-assisted electron tunneling processes in monolayer graphene - hBN - monolayer graphene tunnel transistors.  They allow us to probe electron-phonon interactions in this type of device and identify spectroscopically the energies and nature of the emitted phonons.   Our results suggest that slight misorientations of the component crystalline lattices of these vdW heterostructures may result in differences in the energies and intensities of the measured phonon-assisted tunnel transitions.   

\begin{acknowledgments}
This work was supported by the EU FP7 Graphene Flagship Project 604391, ERC Synergy Grant, Hetero2D, EPSRC (Towards Engineering Grand Challenges and Fellowship programs), the Royal Society, US Army Research Office, US Navy Research Office and US Airforce Research Office.  M.T.G. acknowledges The Leverhulme Trust for support of an Early Career Fellowship. S.V.M. and E.E.V. were supported by NUST ”MISiS” (grant K1-2015-046) and RFBR (15-02-01221 and 14-02-00792). G.J.S. and M.I.K. acknowledges financial support from the ERC Advanced Grant No. 338957 FEMTO/NANO. We are grateful to Gilles de Wijs and Annalisa Fasolino for useful discussions.
\end{acknowledgments}

\end{document}